# Teaching to Learn: iPads as Tools for Transforming Physics Student Roles

Susan Nicholson-Dykstra\*, Ben Van Dusen† and Valerie Otero†

*\*Northglenn High School, Northglenn, CO 80260*
*†School of Education, University of Colorado, Boulder, 80309, USA*

**Abstract.** Students who serve as Learning Assistants (LAs) and have the opportunity to teach the content they are learning, while also studying effective teaching pedagogy, have demonstrated achievement gains in advanced content courses and positive shifts in attitudes about learning science [V. Otero, S. Pollock & N. Finkelstein, Amer J Physics **78**, 11 (2010)]. Although the LA experience is also valuable for high school students, the tight schedule and credit requirements of advanced high school students limit opportunities for implementing traditional LA programs at the high school level. In order to provide high school physics students with an LA-like experience, iPads were used as tools for students to synthesize "screencast" video tutorials for students to access, review and evaluate. The iPads were utilized in a 1:1 tablet:student environment throughout the course of an entire school year. This research investigates the impact of a 1:1 iPad environment and the use of iPads to create "teaching-to-learn" (TtL) experiences on student agency and attitudes toward learning science. Project funded by NSF grant # DUE 934921.

**Keywords:** Teaching to learn, learning assistant, physics education, tablet technology, iPad
**PACS:** 01.40.ekz, 01.40.Fk, 01.50.ht

## INTRODUCTION

At the university level, Learning Assistant (LA) Programs hire talented undergraduates as facilitators to lead small group tutorials that accompany large lecture courses that the LA had previously completed. LA's participate in a three-pronged "Teaching to Learn" (TtL) experience in which they (1) participate in weekly planning sessions to further develop content understanding, (2) attend a weekly education seminar about teaching pedagogy, and (3) practice teaching by leading learning teams.[1] LAs have demonstrated greater performance on conceptual assessments than peers in introductory and advanced content courses. [2,3]

In order to create a similar opportunity for high school students, we previously implemented a modified LA program in four conceptual physics courses in an urban, high-needs school. The high school LAs learned about backwards design, developed lesson plans to teach physics content, then implemented their lessons in a local elementary classroom, reflecting after each experience on the growth of their students.[4] While the experience yielded a clear shift in the high school LAs metacognition and an increased sense of agency, the experience was difficult to implement on a broader scale due to the time requirements for pedagogical training, scheduling requirements, and funding for transportation. University LAs are typically hired to assist in a class they have previously taken. Talented high school students have strict class schedules and minimal time available for facilitating learning environments, making it problematic to implement a traditional LA model in the high schools.

In order to navigate these obstacles, the investigators turned to technology as a tool for developing a modified, asynchronous TtL experience. With the explosion of tablet technology and digital eResources available to students, instructors increasingly face the challenge of purposefully integrating technology into learning experiences. iPads can be used to create "screencast" tutorials in which an iPad app simultaneously records both a student's voice and his/her writing on an iPad screen to create a video tutorial. Tutorials can be assembled into digital libraries of resources for fellow and future students in the course to access, review and evaluate. Students in an AP physics classroom who were piloting a 1:1 iPad environment were challenged with the task of (1) learning required physics content, then (2) developing tutorials to teach peers about the content they had just learned, while (3) also working as a class to develop meaningful standards for "effective screencasts" (aka, digital tutorials or mini-lessons). This LA-like TtL experience better accommodates the scheduling and transportation limitations of high school students.

Because a 1:1 iPad environment is a novel learning situation for most students, the investigators evaluated both the impact of the tablet-rich environment, as well as the impact of the TtL experience, on student understanding of the teaching and learning process, and student opinions of their role in that process.

## METHOD

**Study Context.** This study was conducted in the AP Physics B course in an urban high school during the 2012-13 school year. Of the twenty-six students enrolled in AP Physics, 50% were from ethnic groups traditionally underrepresented in STEM, 19% were female and 15% were English Language Learners. The teacher (first author) is a Noyce Master Teacher in the Streamline to Mastery Program at CU Boulder.

The teacher established a "1:1" iPad environment by providing each student with his/her own device to use during class. The iPads were utilized daily for a variety of activities, ranging from data collection to creation of digital models.

The traditional TtL model for LAs involves three key features: "content," "practice," and "pedagogy". [1] All students in the study were enrolled in AP Physics, studying traditional physics content. To provide an opportunity for students to "practice" teaching content, students utilized iPads to create "screencasts" in which they taught a mini-lesson on how to solve a homework problem. Student-generated screencasts were assembled into a database to serve as an eResource for other physics students (peers in class and future AP students). In order to integrate teaching pedagogy into the experience, students were asked to watch videos created by their peers, provide feedback, and formulate a list of criteria for successful lesson presentations. These criteria were assembled into a rubric for guiding screencast creation. Students periodically revised the rubric based on discussions about the goal of the assignment.

**Data Collection and Analysis.** Two types of data were collected: student surveys and investigator observations. Student surveys were designed for students' reflection on their experiences using iPads and were primarily composed of open-ended reflection and Likert-style questions. For statistical analysis of Likert-style questions, "Strongly agree" and "agree" responses were combined. Open-ended reflection responses were coded inductively and categorized. The frequency of occurrence of each category was determined and is presented as a percentage. Some statements provided evidence for multiple categories, so percentages do not add up to 100%.

Classroom reflections were documented by the teacher, specifically noting the types of formal and informal interactions students had around iPads use. These qualitative findings are central to observing the shift in student perspectives on learning and their role in the classroom. Examples of student screencasts and video reflections about iPads and screencasts are available online (http://tinyurl.com/nes5u4g).

## RESULTS

**Finding 1: The 1:1 iPad environment personalizes learning experiences**

Students were surveyed at the end of the year regarding their opinion of how a 1:1 iPad environment had impacted their learning (Table 1). The results indicate that the students' use of technology personalized their learning by providing a more self-guided experience. Students felt that iPads transformed their roles from student to being more like an independent researcher. Students reported that they were allowed to work independently while self-pacing and seeking answers to their personal questions throughout class. Additionally, all students indicated that they modified their assigned iPad in some manner in order to better fit their learning needs or preferences (e.g. modifying background image, sorting apps, modifying keyboard organization/language, or providing maintenance), showing personalization of their learning tool. Table 2 shows that a significant percentage of students felt that iPads made learning more fun and engaging, and helped them to learn science. Students also indicated that the iPad served as a tool for increasing their communication with both their instructor and their peers.

**TABLE 1.** Survey on iPad's impact on learning (N=14)

| iPads allow me to: | |
|---|---|
| Work independently | 79% |
| Research my own questions | 79% |
| Communicate more with my teacher | 71% |
| Work at my own pace | 64% |
| Be more creative in my work | 64% |
| Communicate more with my peers | 57% |
| I modify (personalize) my iPad in at least one way to meet my needs & preferences | 100% |

**TABLE 2.** Survey on how iPads help students learn (N=14)

| iPads help me to learn science… | 71% |
|---|---|
| By supporting my personal learning needs (pace, type of activity, ease of use) | 49% |
| By providing access to resources (internet, apps, tools) and allowing me to be the researcher | 44% |

Students were also asked to explain how iPads helped them to learn. From the responses, two primary categories emerged. The first category was, "iPads provide access to resources and allow student-directed learning and research". For example, *"[iPads] allow me to use the internet to research any questions I have and allows us to communicate between ourselves*

*(students)"*. The second category was, "iPads support personal needs to optimize learning". For example: *"[iPads provide] the freedom to work through assignments without waiting on or hindering the class";* and *"[iPads] allow me to look at problem solving visually, which is the best way I learn"*.

**Finding 2: Student participants in the TtL demonstrate greater metacognition regarding the teaching and learning process.**

*I. Participants demonstrate concern for student engagement and growth*. For screencast assignments, students were asked to "create a tutorial teaching how to solve a problem." With the exception of a single example at the start of the year, students were not provided with any guidelines or requirements, and were asked to develop their own strategy for creating screencasts. After each screencast assignment, students were asked to watch, reflect on, and provide feedback about videos created by their peers, and periodically, the class discussed what made a good screencast. In a particular situation, prior to the first class discussion, the investigator observed students commenting on one student's product. Comments included responses such as "It's hilarious!", "I wonder what he'll do next time", "That made me think…", and "I think I'll try that next time…". These comments indicated not only student excitement for reviewing and listening to peer products, but also for modifying their own instructional strategy. What had this student done to make his screencast so remarkable? He taught his entire lesson using a foreign accent. Students found his modification entertaining and engaging, and many emulated his work in future screencasts.

After completing four screencasts, students were required to submit a list of characteristics of effective screencasts. During a subsequent class discussion, students compiled a comprehensive list of key criteria, then worked in groups to sort these into three categories (content, organization, and aesthetics), and into levels of instructional proficiency (proficient vs. advanced). One of the characteristics that students believed made a lesson exceptional was the inclusion of humor. One common student concern that arose was how to engage (and entertain) an audience if a student was not comfortable using humor or accents. Based on this concern, students revised their rubric to include humor as one method of engaging an audience.

Additionally, students worked with their peers to brainstorm other ways to engage an audience without the use of humor, but also set aside time (outside of class) to playfully practice an accent or a joke that they could include in their screencast if they wanted to try out that strategy. One student was determined to develop her own twist on humor and accents. After completing her last screencast, she proudly requested the opportunity to share her lesson with the class. This student had enlisted a classmate to talk in a "Darth Vader" accent to begin the lesson, then she cleared her throat and continued her lesson in her own voice.

Through discussion of effective lesson delivery, students recognized the necessity of delivering content, engaging their audience, and providing a clear and correct lesson. These discussions provided opportunities to discuss the idea of "best practice" and "effective teaching pedagogy", transitioning students from the role of learner to teacher.

*II. Participants induce principles of best practice.* At the conclusion of semester 1, students had completed 4 screencasts and were asked to explain why they believed they were required to create screencasts. Based on open-ended responses, 5 codes emerged regarding student perspectives on the value or usefulness of screencasting as a learning activity:

1. <u>Teaching to Learn</u>: Students reported that the process of teaching a lesson helps them to develop agency, specifically a deeper understanding of the content and problem-solving process. For example, *"[Screencasting] helped me understand physics better because while doing the activity, it gives us a chance to practice what we've just learned in class. This enables us to hammer the content into my head."* Another example, *"[Screencasting] helps me to understand topics via explaining them out loud and working them out"*.

2. <u>Learn from peers</u>: Students reported that the opportunity to observe multiple perspectives and processes helps them to better understand the content. For example,*"[Screencasting] allows the students to see how others are solving similar problems"* and, *"We get a different person's view on each subject"*.

3. <u>Evaluate Understanding</u>: Students reported that screencasts help them to evaluate and identify gaps in their content understanding, while also providing evidence to their instructor of their progress. For example, *"[Our teacher] can see the way we do problems and see if and where we made a mistake"* and, *"If you can teach it to someone then you understand it more and it shows you how much you actually understand."*

4. <u>Practice Communicating Science</u>: Students reported that screencasting helps them to practice communicating their understanding and developing skills for delivering a clear, scientifically correct lesson. For example, *"[Screencasting helps] to be able to explain our thought process"* and, *"We can go step by step explaining things to other classmates"*.

5. <u>Supporting and Helping Peers</u>**:** Students reported feeling empowered by the process to help support their peers' academic growth. All other categories were focused on the growth and practices of the student engaged in TtL, but this category focuses on the altruistic impact on their audience. For example, *"[We] create screencasts to reinforce our knowledge on the subject we are doing the screencast on while also helping any fellow classmates who might need help on the topic"* and *"we are able to help one another understand."*

**TABLE 3.** Student reflections on screencast value (n=22)

| | |
|---|---|
| Teaching to Learn | 65% |
| Learn from Peers | 46% |
| Practice Communicating Science | 15% |
| Evaluate Understanding | 10% |
| Support/Help Peers | 10% |

Table 3 indicates the percentage of student responses that aligned with each of the five dominant themes. Based on the emergent themes, students indicated that teaching to learn not only pushes them to learn through the experience of teaching or preparing to teach, but also provides opportunities to (asynchronously) learn from their peers. Student responses indicate a concern for personal ability to communicate effectively and assess understanding, as well as a concern for the growth of their peer learners. These foci align closely with those of a growth-minded teacher.

*III. Participants compare impact of different learning activities.* Near the end of semester 2, students had completed 8 screencasts and were asked to identify and explain the differences between screencasts and traditional homework activities (Table 4). Most responses suggested that screencasts require a greater depth of content understanding and grasp of problem solving strategies.

Students reported that screencasting requires more communication, allows for more creativity and freedom of expression, and engages multiple senses, all ideas involving attention to audience engagement. Students also indicated that screencasting involves an audience, which serves to impose a different level of accountability for the "teacher" to learn the content material. In contrast, students apparently do not identify an audience for traditional homework and therefore feel less vested in understanding the content.

## CONCLUSIONS

The TtL experience that the AP physics students participated in was different from the traditional LA model—students had not already been exposed to the content, students were not able to interact with their mentees for Socratic questioning and assessment of audience understanding as occurs in traditional Peer Instruction, [1-3] and there was never explicit discussion of best practices in physics education or teaching pedagogy. Yet, students exhibited greater metacognition regarding the teaching and learning process, with a focus on student (audience) engagement, assessment and growth. Students also identified their roles in the learning process as being both learner and teacher. Although the structure of the teaching to learn experience differed significantly from the traditional LA model, students exhibited similar sophistication of understanding of learning.

Ongoing studies are evaluating the impact of this modified TtL experience on content mastery and students' perspectives of their roles as teachers and researchers. Future studies will investigate strategies for making the screencasting experience more interactive for the teacher and learner.

**TABLE 4.** Student-Designated Differences Between Screencasts and Traditional Homework (n=19)

| | |
|---|---|
| Screencasts requires more work & greater depth of understanding | 53% |
| Screencasts require teaching the process | 37% |
| Screencasts provide freedom for more creative expression | 37% |
| Screencasts allow for more communication | 32% |
| Screencasts have a specific audience and allow the audience to learn the process | 26% |
| Screencasts impose a different level of accountability and motivation for understanding content | 21% |
| Making and watching screencasts is a multi-sensory learning experience | 16% |
| More intimidating to seek help for screencasts | 11% |
| Traditional homework requires greater breadth of understanding | 5% |
| No significant differences | 5% |